\documentclass[3p]{elsarticle}

\usepackage{color}
\usepackage{url}
\usepackage{float}
\usepackage[utf8x]{inputenc}

\journal{arXiv}

\begin{document}

\begin{frontmatter}

\title{Robustness analysis in an inter-cities mobility network: modeling municipal, state and federal initiatives as failures and attacks}


\author[UFOP]{Vander L. S.  Freitas\corref{mycorrespondingauthor}}
\cortext[mycorrespondingauthor]{Corresponding author}
\ead{vandercomp@gmail.com}

\author[UNESP]{Jeferson Feitosa}
\author[Cemaden]{Catia S. N. Sepetauskas}
\author[Cemaden,HUB]{Leonardo B. L. Santos}

\address[UFOP]{Department of Computing, Federal University of Ouro Preto (UFOP), Ouro Preto, Brazil}
\address[UNESP]{São Paulo State University (UNESP), Sao Jose dos Campos, Brazil}
\address[Cemaden]{National Center for Monitoring and Early Warning of Natural Disasters (Cemaden), Sao Jose dos Campos, Brazil}
\address[HUB]{Institute of Physics, Humboldt University of Berlin, Germany}

\begin{abstract}

Motivated by the challenge related to the COVID-19 epidemic and the seek for optimal containment strategies, we present a robustness analysis into an inter-cities mobility complex network. We abstract municipal initiatives as nodes' failures and the federal actions as targeted attacks. The geo(graphs) approach is applied to visualize the geographical graph and produce maps of topological indexes, such as degree and vulnerability. A Brazilian data of 2016 is considered a case study, with more than five thousand cities and twenty-seven states. Based on the Network Robustness index, we show that the most efficient attack strategy shifts from a topological degree-based, for the all cities network, to a topological vulnerability-based, for a network considering the Brazilian States as nodes. Moreover, our results reveal that individual municipalities' actions do not cause a high impact on mobility restrain since they tend to be punctual and disconnected to the country scene as a whole. Oppositely, the coordinated isolation of specific cities is key to detach entire network areas and thus prevent a spreading process to prevail.

\end{abstract}


\end{frontmatter}


\section{Introduction}

The complex network approach \cite{NetScienceEEStructure} emerges as a natural mechanism to handle mobility data, taking areas as nodes and movements between origins and destinations as edges \cite{Barbosa2018, Santos2019_TTC}. The structure of the underlying network of a system reveals its ability to survive to failures and coordinated attacks. One important question is to know how many nodes can be removed until the network completely fragments into small pieces \cite{NetScienceB}. In this paper, we present a robustness analysis \cite{Callaway_et_al_2000,NetScienceB} on mobility complex networks, motivated by the challenge related to COVID-19 epidemic and the seek for optimal containment strategies.

As of April 6th, 2020, the pandemic of COVID-19 has reached more than 200 countries, with more than 1,280,000 confirmed cases and more than 70,000 deaths, globally. In Brazil, there are more than 11,200 confirmed cases and more than 480 deaths \cite{Coelho2020, site}.

Our analysis offers tools for decision support in multiple scenarios like containment of the ongoing epidemiological spreading \cite{Santos2009, epi}, the detection of vulnerable areas for disasters \cite{Madeira, NHESS}, or cities whose importance in mobility is vital for the transit of people and supplies. In the former case, there is also a trade-off between interrupting the spread through isolation and losing a crucial route. 

Within the context of the robustness analysis, the local/municipal initiatives are modeled as failures and the federal's as attacks. Our results consist of strategies to effectively damage the network structure by choosing the cities (or states) that have more impact in mobility. The local initiatives are considered failures because they are sometimes disconnected from the federal ones. It is possible that some cities start to care about an epidemics before the others, and/or before the country itself, either because their mayors have more political influence than the average, or due to local popular pressure. In both cases, the outcome for the city is likely to diverge from the announced measures for the country.

Contrarily, the coordinated attacks in the mobility network are considered to be a federal action due to its global scale characteristics. As the entire network is subjected to the federal rules, the state has the power to pick specific nodes, either to invest in infrastructure because of its potential flow for supplies, to diminish natural disaster risks, or to isolate them from the rest in a disease outbreak.

The most commonly used mobility data for such analysis in Brazil is the shorter distance pendular travels, from the 2010 national census (IBGE) \cite{Pendular}. In this paper, we use the roads IBGE data from 2016 \cite{mobIBGE}, which presents the flow of buses, vans and similar transports between cities, considering only vehicles from companies that sell tickets to passengers. The information collected by this research seeks to quantify the interconnection between cities, the movement of attraction that urban centers carry out for the consumption of goods and services and the long-distance connectivity of Brazilian cities.

Our contributions are the robustness analysis of the Brazilian inter-cities mobility network with the abstraction of nodes' failures as municipalities' actions and targeted attacks as federal's. We assess the network impacts during nodes' removal through a robustness index, which is computed 
from the giant component size.
Moreover, we present maps of the cities colored according to their respective control indexes that guide the coordinated attacks, and consequently, exhibit which cities play a key role in mobility restrain.

This paper is organized as follows: Section \ref{sec:method} presents the data and the methods we employ, such as the vulnerability index calculation, the network robustness measure, and the geographical visualization tools. The robustness analysis is performed in Section \ref{sec:robustness}, followed by the results and discussion in Section \ref{sec:results_discussion}. Lastly, the final remarks are depicted in Section \ref{sec:final_remarks}.

\section{Method}
\label{sec:method}

The IBGE data \cite{mobIBGE} contains the weekly travel frequency (flow) between pairs of Brazilian cities/districts by transports in which it is possible to buy a ticket, mainly buses and vans, but also aquatic transports. The frequencies are aggregated within the round trip, which means that the number of travels from city A to city B is the same as from B to A. We produce three types of undirected networks with a different number $N$ of nodes to capture actions in different scales (country and state):
\begin{enumerate}
    \item $N=5420$ - Brazil\footnote{Almost all brazilian cities are considered in the dataset - details in [13]} (BR): nodes are cities and edges are the flow of direct travels between them.
    \item $N=620$ - São Paulo state (SP): a subset of the previous network, containing only cities within the São Paulo state.
    \item $N=27$ - Brazilian states (BS): in contrast with the others, here each state is a node and the edges are the accumulated flows between them.
\end{enumerate}

Several networks are analyzed from the three models (BR, SP, and BS). We employ certain thresholds to neglect travels with three levels of frequency and build the following networks: i) considering all recorded flow, ii) with edges of at least an average flow and, iii) a more restricted topology with only the higher flows. The chosen thresholds are $\eta_1 = 0$, $\eta_2 = \mu$ and $\eta_3 = \mu+\sigma$, in which $\mu$ is the mean recorded flow for the network and $\sigma$ is the standard deviation. We end up with nine networks in total as described in Table \ref{tab:networks}. The motivation behind the threshold levels is based on the fact that people avoid traveling during epidemics and companies may close routes with fewer people volume.

\begin{table}[]
\centering
\caption{ \label{tab:networks} Statistics for the Brazilian (BR), São Paulo state (SP) and Brazilian states (BS) networks, with three flow thresholds: $\eta_1=0$, $\eta_2=\mu$ and $\eta_3=\mu + \sigma$, where $\mu$ is the average flow and $\sigma$ is the standard deviation. The average degree $\left \langle k \right \rangle$ is the average number of connections the nodes have, $|E|$ is the number of edges in the network, and $N$ is the number of nodes.}
\vspace{0.1cm}
\begin{tabular}{llll|l|l|l|l|l|l|}
\cline{5-10}
                              &                              &                               &      & \multicolumn{2}{l|}{$\eta_1$}         & \multicolumn{2}{l|}{$\eta_2$}         & \multicolumn{2}{l|}{$\eta_3$}         \\ \hline
\multicolumn{1}{|l|}{Network} & \multicolumn{1}{l|}{$\mu$}   & \multicolumn{1}{l|}{$\sigma$} & $N$  & $|E|$ & $\left \langle k \right \rangle$ & $|E|$ & $\left \langle k \right \rangle$ & $|E|$ & $\left \langle k \right \rangle$ \\ \hline
\multicolumn{1}{|l|}{BR}      & \multicolumn{1}{l|}{48.04}   & \multicolumn{1}{l|}{100.21}   & 5420 & 65264 & 24.08                            & 15505 & 5.72                             & 4217  & 1.56                             \\ \hline
\multicolumn{1}{|l|}{SP}      & \multicolumn{1}{l|}{73.20}   & \multicolumn{1}{l|}{122.79}   & 620  & 9592  & 30.94                            & 2610  & 8.42                             & 758   & 2.45                             \\ \hline
\multicolumn{1}{|l|}{BS}      & \multicolumn{1}{l|}{2032.29} & \multicolumn{1}{l|}{4397.86}  & 27   & 474   & 35.11                            & 108   & 8.00                             & 44    & 3.25                             \\ \hline
\end{tabular}
\end{table}

\vspace{2cm}

\subsection{Vulnerability index}
\label{sec:vuln}

This section presents a pointwise index related to the network and associated with each node, the so-called Vulnerability index \cite{LivroSpringer, NHESS}:

\begin{equation}
    V_k = \frac{ E - E_k^* }{E},
\end{equation}

\noindent with $E$ being the network Efficiency:

\begin{equation}
    E = \frac{\sum_{i,j \in V, i \neq j} e_{ij}}{N(N-1)},
\end{equation}

\noindent where the efficiency $e_{ij}$ in the communication between nodes $i$ and $j$ is defined as inversely proportional to the shortest path length between them, and $E_k^*$ is the network efficiency after the removal of the $k$-th node. In short, networks with small shortest path lengths are more efficient. The  $V_k$ quantifies how vulnerable a network is when a certain node is deleted.


\subsection{Robustness}
\label{sec:robustness}

The robustness of a network is its capacity of keeping connected even after the removal of nodes or edges. This process may be performed in a random fashion such as an energy drop reaching some computers in computer networks, or a car accident on an important road, in a mobility network. Those events are usually not predictable and depend on several internal and/or external causes, thus characterizing a system failure. Conversely, a node may be intentionally removed in order to disrupt the network structure, typifying an attack.

We propose strategies to identify the municipalities that play a key role in mobility. First, the network response to random failure is assessed as a baseline to compare with attacks to nodes of either high degree or whose removal enhances the network vulnerability. Our motivation is the fact that real networks are robust to random failures but are fragile to attacks \cite{NetScienceB,Callaway_et_al_2000,Cohen_et_al_2000,Iyer_et_al_2013}. 

The main question is to figure out how many and which nodes must be removed until the network collapses. This being said, understanding which cities are important for mobility to either invest in infrastructure to enhance their capacity or to know exactly which node should be isolated in a disease outbreak is of major interest.

The measure we use to quantify the network response to both failures and attacks is the number of nodes in the giant component $P_\infty(f)$,  when a certain rate $f$ of nodes is removed, which captures whether the network remains connected or not.

Choosing the proper node to be removed is crucial and can be done in different ways. Failures are the trivial case, for which a node is randomly selected. However, coordinated attacks demand some strategy like always removing the nodes with higher degrees. We propose two strategies: deleting nodes with higher degree ($\max(k)$) or higher $V_k$ ($\max(V_k)$). Attacks oriented by higher degrees are effective and produce better results than non-local measures in most cases \cite{Iyer_et_al_2013}. The $V_k$ is also applied here due to its direct relation to mobility since it captures how vulnerable the network becomes in the absence of certain nodes.



The BR network ($N=5420$) has a degree distribution that follows a power-law with coefficient of $\gamma=2.57$, which characterizes a scale-free topology. This means that, under random failures, the critical threshold for which the network is broken is around $f_c=0.9911$, for $f_c = 1 - ( 1 / (\kappa - 1))$ with $\kappa = \left \langle k^2 \right \rangle / \left \langle k \right \rangle$. This represents a structure that is strongly robust to failures, i.e., almost all nodes must be removed before the giant component is dismantled \cite{NetScienceB}. 

On the other hand, such networks are vulnerable to attacks, especially when they are targeted to higher degree nodes (hubs). Within this context, we consider random failures as isolated mitigation actions by some city or state and attacks as federal mitigation actions (based on the ``big picture''). Regarding mobility networks, some cities are key to an outbreak disruption. Consider for instance the isolation of Wuhan in the restrain of COVID-19 spread in China \cite{Li_et_al_2020}. Isolating São Paulo when the first cases appeared could have substantially reduced the number of cases in Brazil.

The robustness is measured by

\begin{equation}
    R = \frac{1}{N} \sum_{i=1}^N \frac{P_\infty(i/N)}{P_\infty(0)},
\end{equation}

\noindent for $R \in (0,1/2)$ and $P_\infty(i/N)$ is the number of nodes in the giant component when a fraction $i/N$ of the nodes are removed. The higher the $R$, the more robust the network is. Note that the normalization factor $1/N$ allows different size networks to be compared to each other. The minimum $R$ value of $1/N$ is reached with the star-like topology and the maximum $\frac{1}{2}(1-1/N)$ with the complete graph.



\subsection{Geographical visualization}
\label{sec:GG}

A geographical approach for complex systems analysis is especially important for mobility phenomena \cite{barthelemy2011spatial}. Santos {\it et al.} (2017) \cite{Santos2017} proposed the (geo)graphs approach, in which a (geo)graph is defined as a graph where the nodes have a known geographical location, and the edges have spatial dependence. They provide a simple tool to manage, represent and analyze geographical complex networks in different domains \cite{Santos2019_TTC, Seron2019} and it is used in the present manuscript. The geographical manipulation is performed into PostgreSQL Database Management System with its spatial extension PostGIS and the maps are produced using the Geographical Information System ArcGIS.

\section{Results and discussion}
\label{sec:results_discussion}

This section presents the results of the robustness analysis for the nine previously mentioned networks. As stated, the nodes are connected when between them there is a nonzero flow, which means that the number of connections decreases for increasing thresholds ($\eta$). 

\begin{figure}[H]
    \centering
    \includegraphics[width=0.9\linewidth]{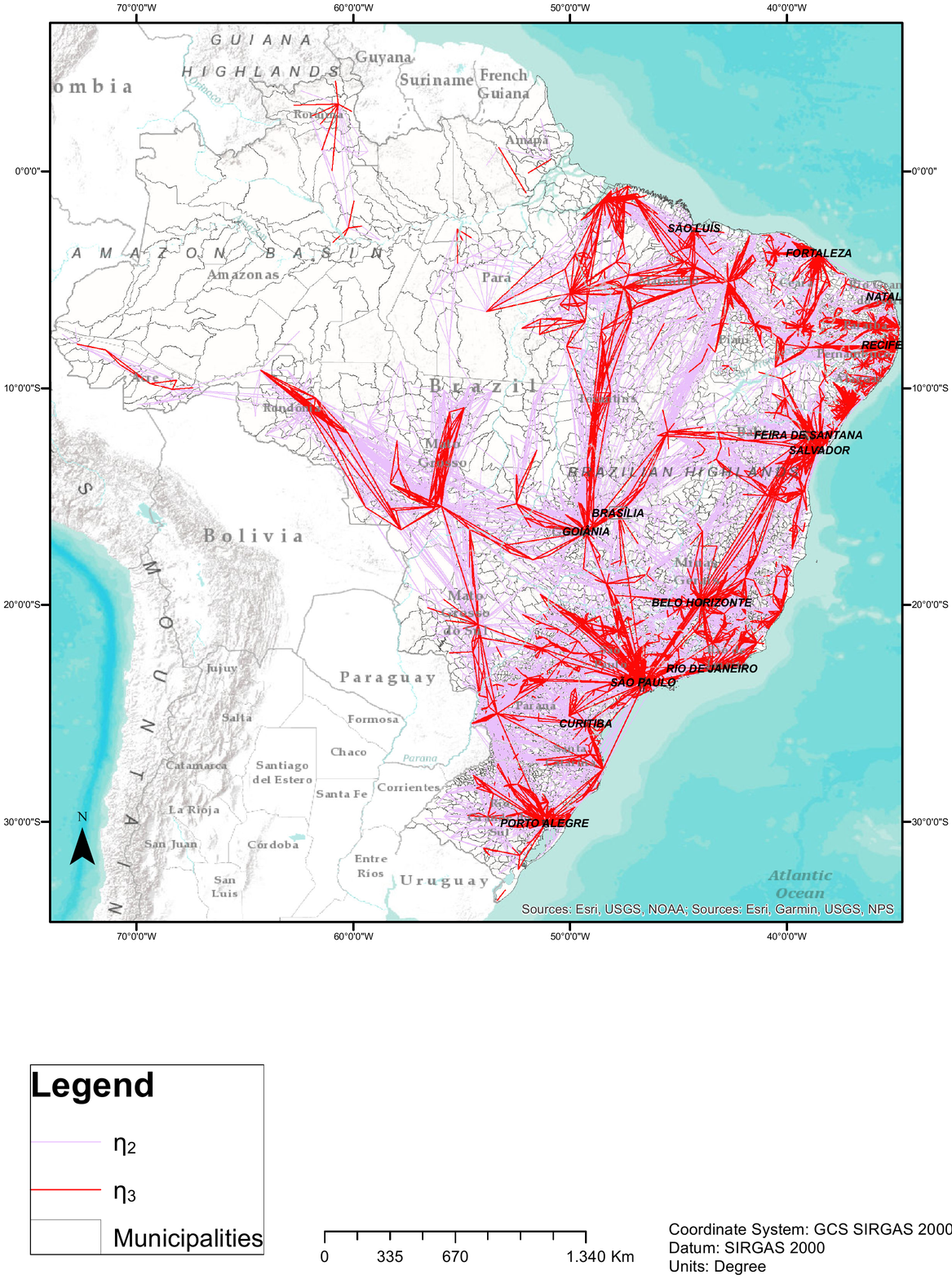}
    \caption{Map with the sets of edges for different values of $\eta$, for the network of Brazilian cities.}
    \label{fig:GG}
\end{figure}

Following the (geo)graphs approach, it is possible to visualize nodes and edges of the Brazilian mobility network in the geographical space (Figure \ref{fig:GG}) for different values of $\eta$. It is important to highlight some key cities like Belo Horizonte, Rio de Janeiro, São Paulo and Salvador, and the large number of connections between them.

\begin{figure}[H]
    \centering
    \includegraphics[width=0.9\linewidth]{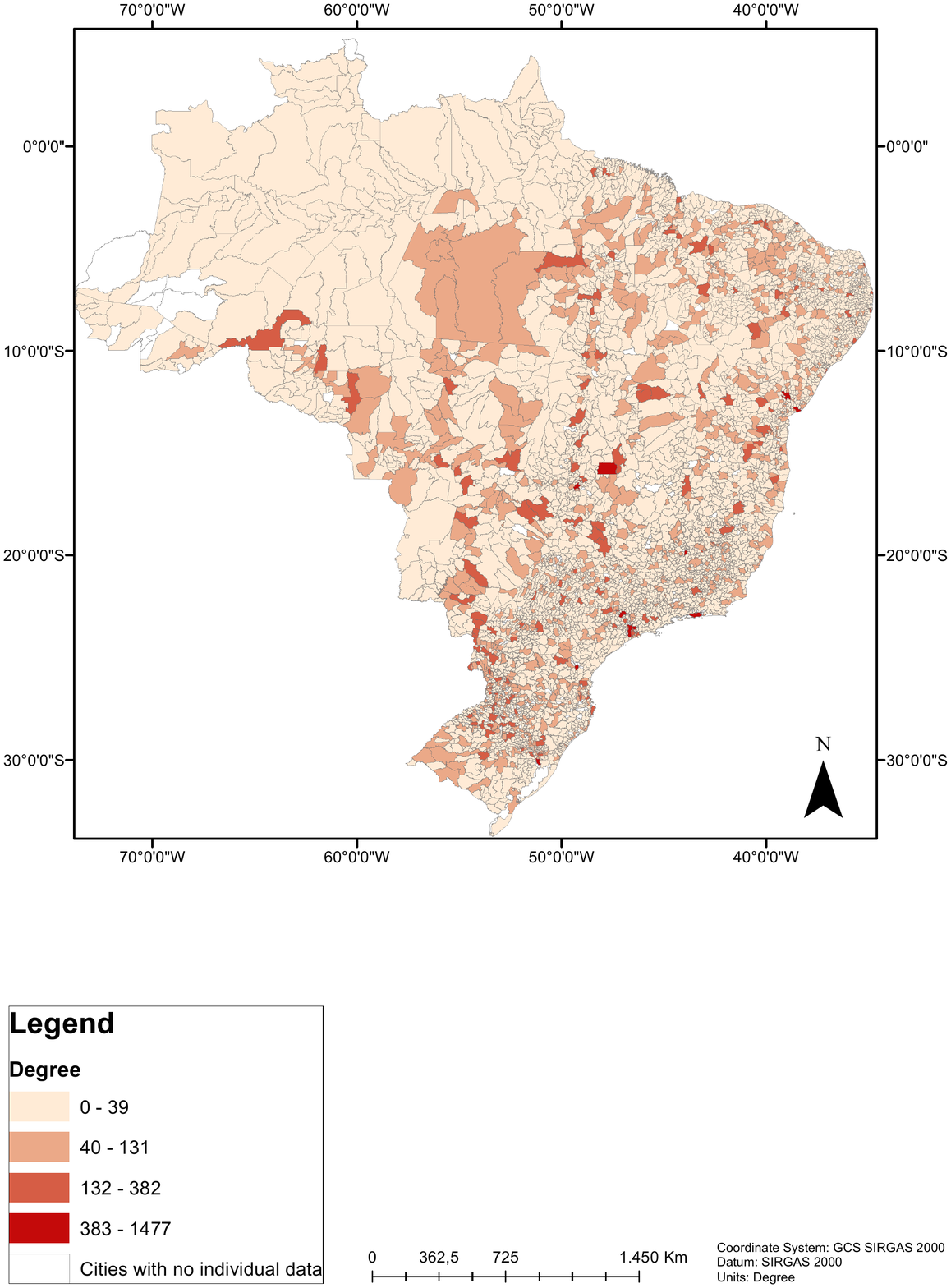}
    \caption{Map of the topological degree index related to each node/city, for $\eta_1=0$.}
    \label{fig:Deg_0}
\end{figure}

\begin{figure}[H]
    \centering
    \includegraphics[width=0.9\linewidth]{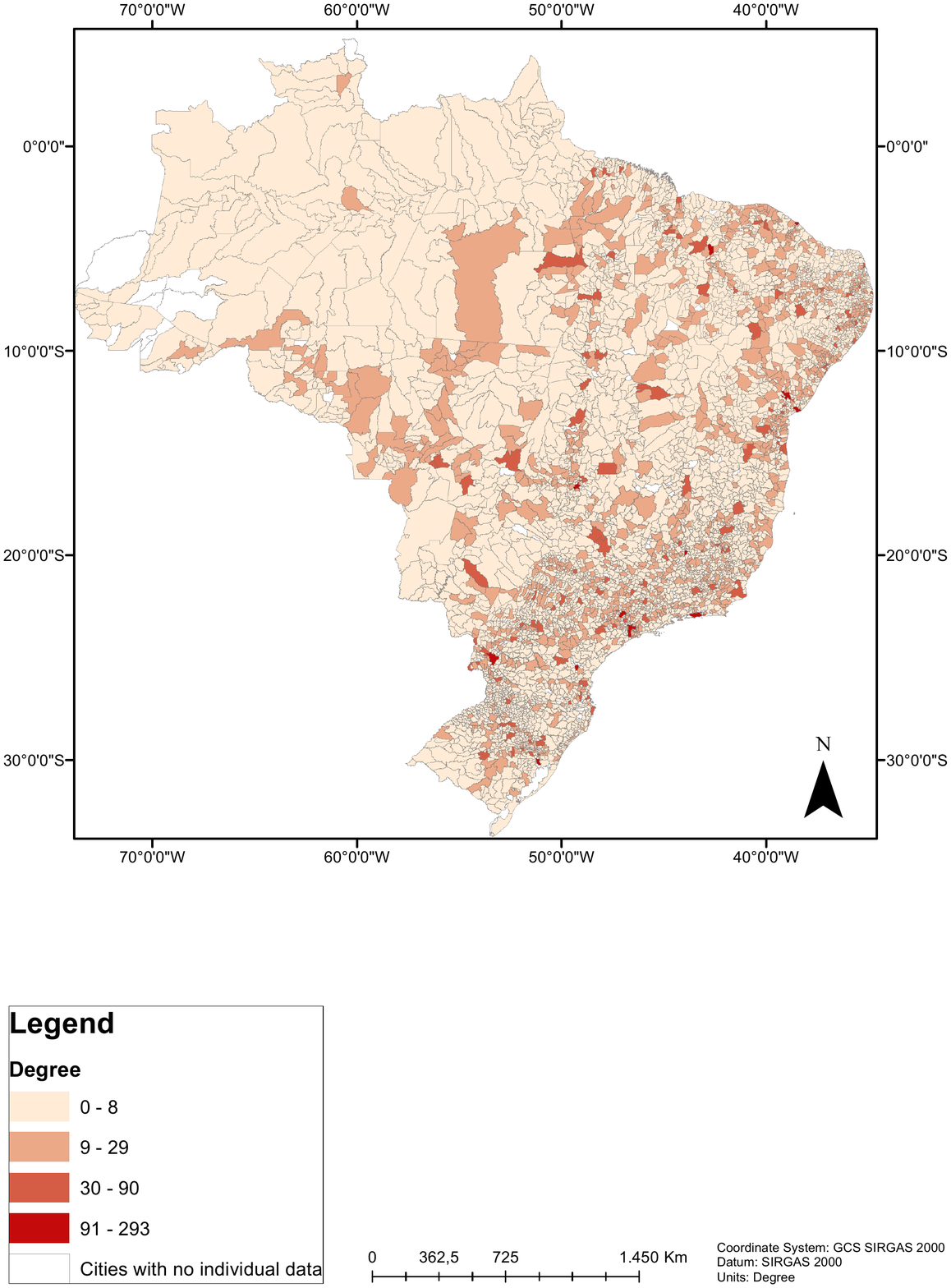}
    \caption{Map of the topological degree index related to each node/city, for $\eta_2=48$.}
    \label{fig:Deg_48}
\end{figure}

\begin{figure}[H]
    \centering
    \includegraphics[width=0.9\linewidth]{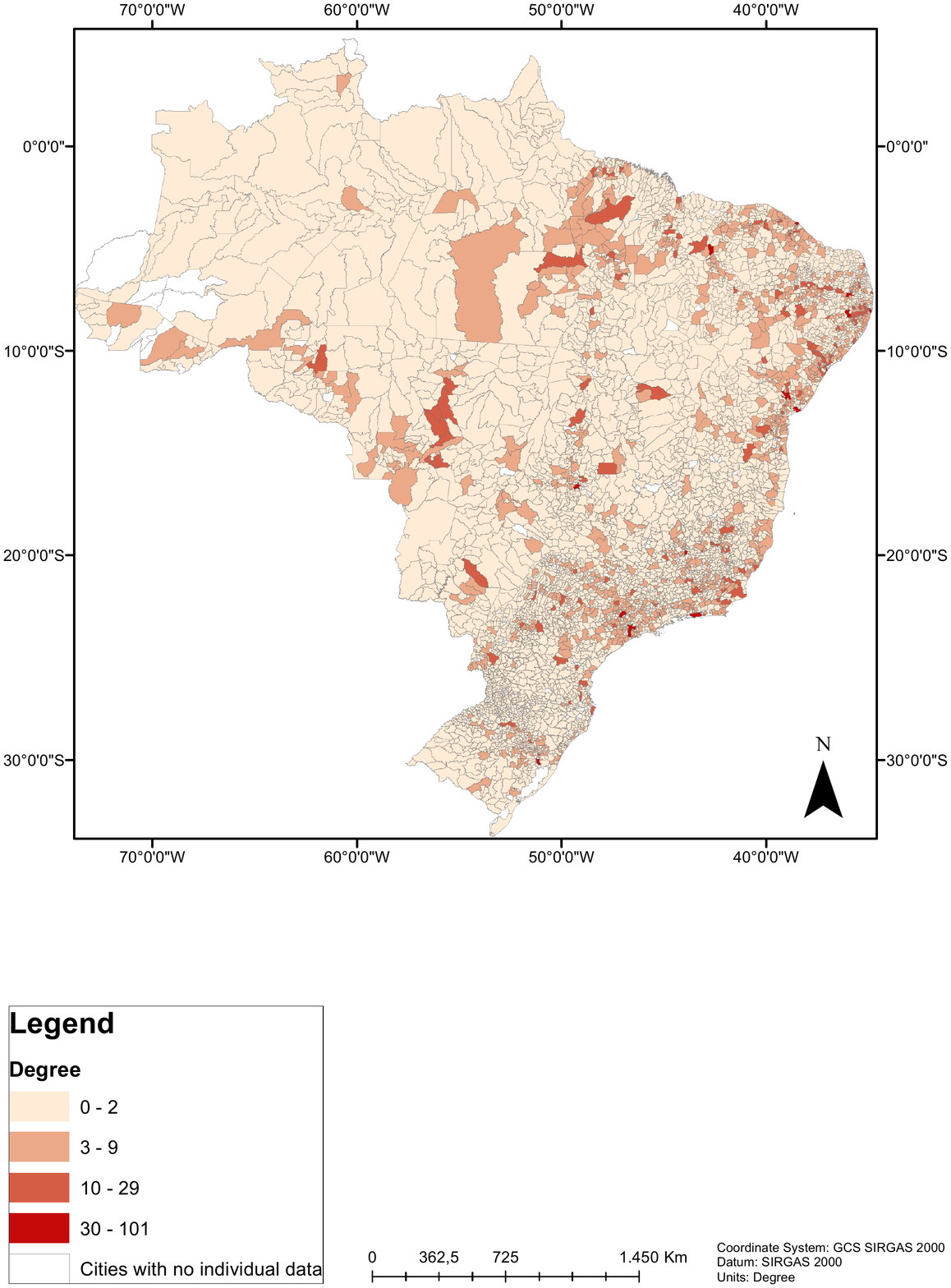}
    \caption{Map of the topological degree index related to each node/city, for $\eta_3=148$.}
    \label{fig:Deg_148}
\end{figure}

\begin{figure}[H]
    \centering
    \includegraphics[width=0.9\linewidth]{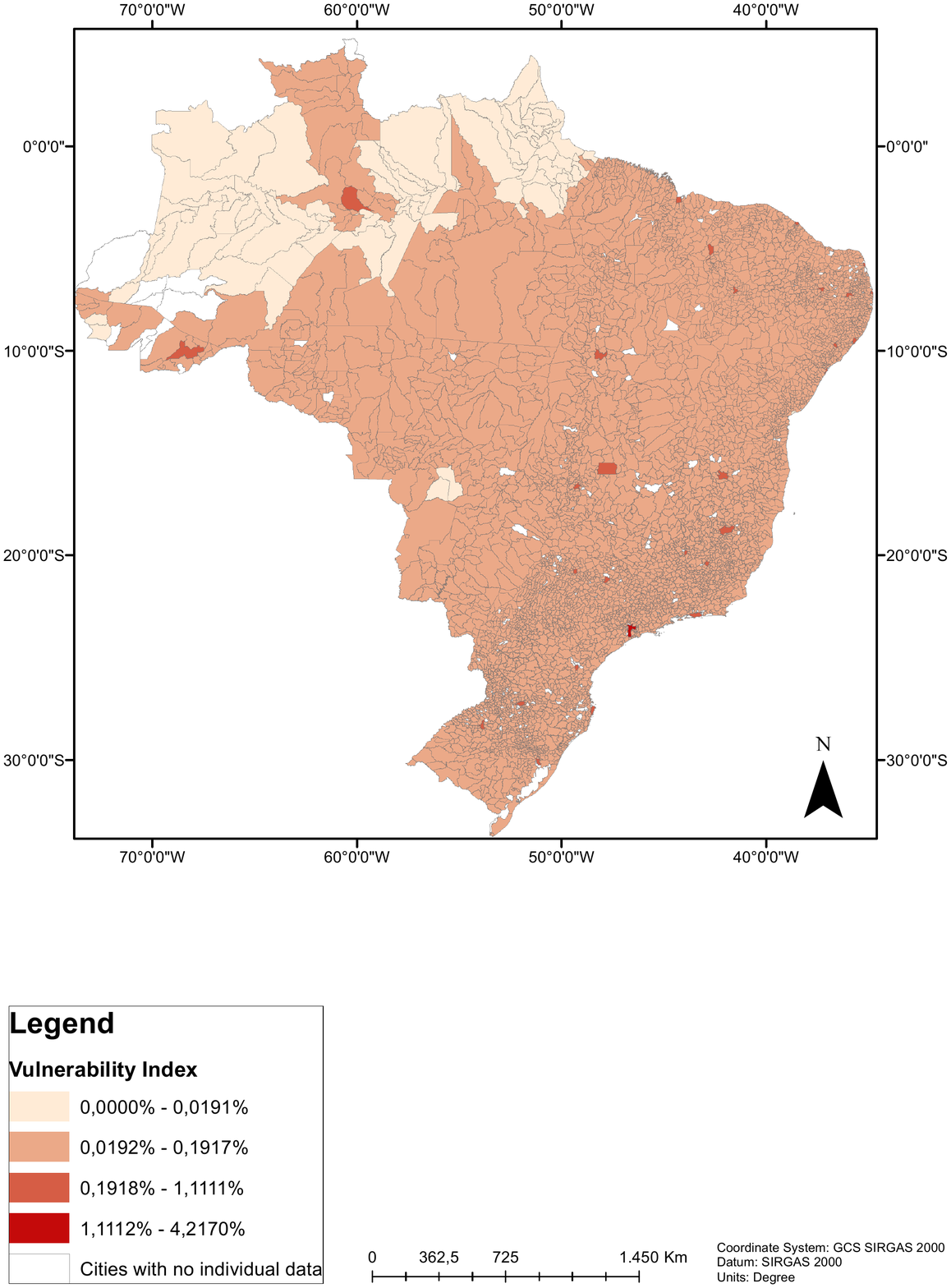}
    \caption{Map of the topological vulnerability index related to each node/city, for $\eta_1=0$.}
    \label{fig:Vul_0}
\end{figure}

\begin{figure}[H]
    \centering
    \includegraphics[width=0.9\linewidth]{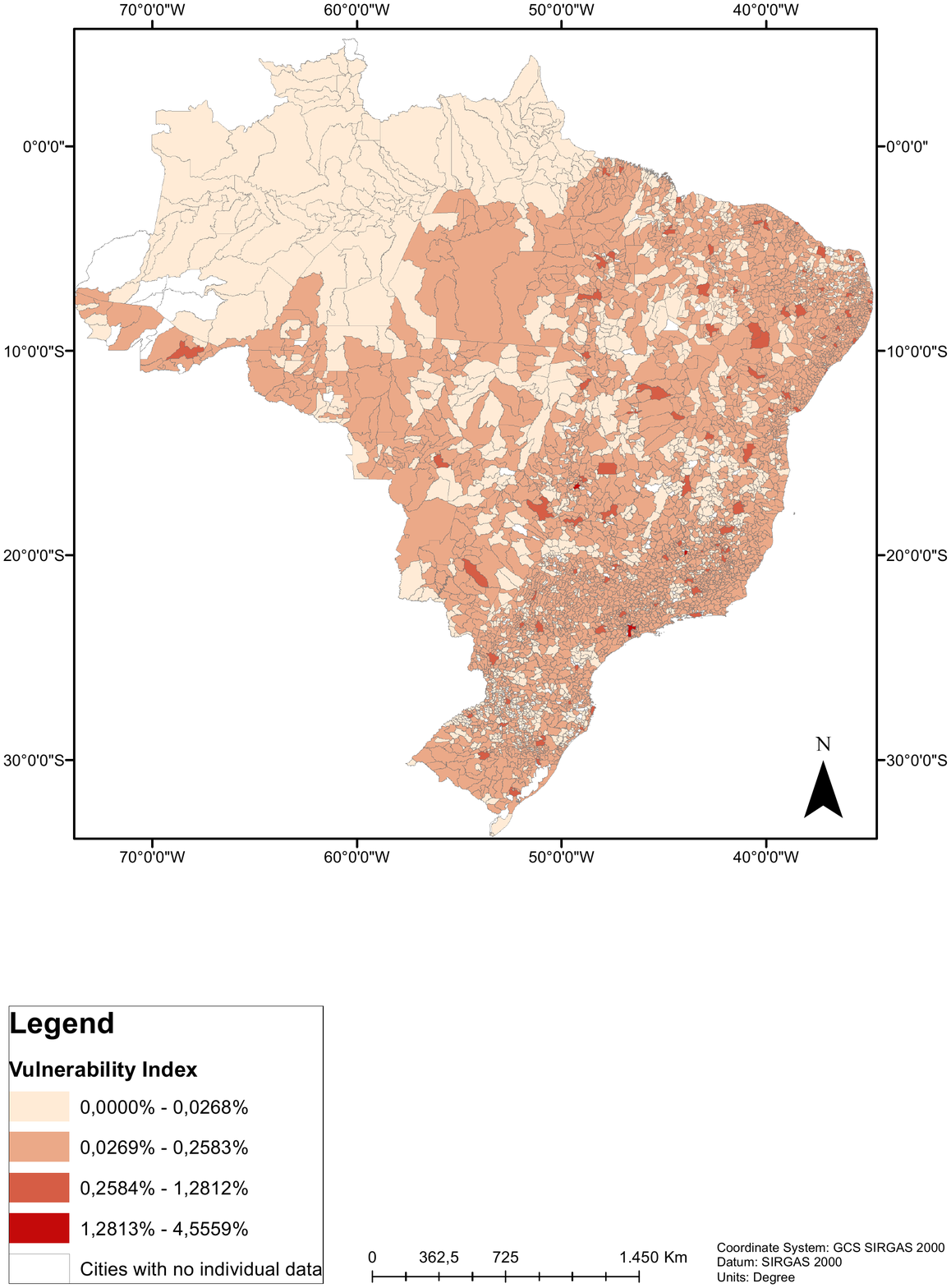}
    \caption{Map of the topological vulnerability index related to each node/city, for $\eta_2=48$.}
    \label{fig:Vul_48}
\end{figure}

\begin{figure}[H]
    \centering
    \includegraphics[width=0.9\linewidth]{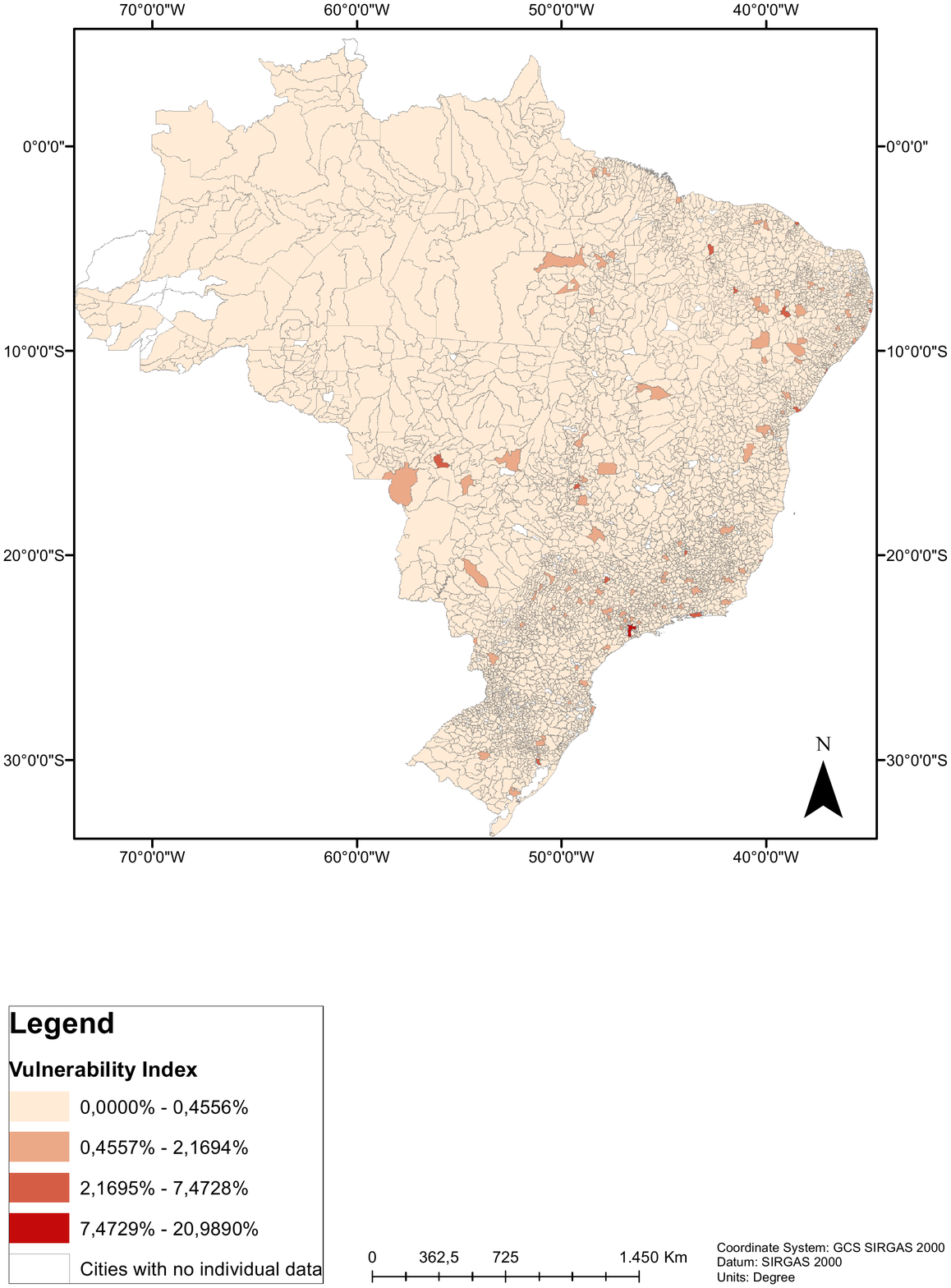}
    \caption{Map of the topological vulnerability index related to each node/city, for $\eta_3=148$.}
    \label{fig:Vul_148}
\end{figure}

Figure \ref{fig:Deg_0} shows the map of the topological degree index related to each node/city, considering all edges ($\eta_1=0$). In Figure \ref{fig:Deg_48} there is the equivalent for $\eta_1=48$ and in Figure \ref{fig:Deg_148} for $\eta_2=148$. Figure \ref{fig:Vul_0} presents the map of the topological vulnerability index related to each node/city for $\eta_1$, Figure \ref{fig:Vul_48} has the equivalent for $\eta_2$, followed by Figure \ref{fig:Vul_148} with $\eta_3$. It is possible to see the high spatial heterogeneity associated to these topological indexes, especially for high threshold values. The Pearson correlation between $k$ and $V_k$ is about $R^2 = 0.5$ for the BR network, showing that they capture similar properties but are still different as we show later with the robustness index.

When the Figures for degree and vulnerability with the same thresholds $\eta$ are compared, it seems like the higher degree nodes are more highlighted, while vulnerability indexes are mostly homogeneous. In fact, the average degree corresponds to 2\% of the maximum observed degree and the average vulnerability is about 1\% or less of the $\max(V_k)$. The representation of $V_k$ would look like $k$ in the absence of $\max(V_k)$ or in a different distribution of color scales.



\begin{figure}[H]
    \centering
    \includegraphics[width=0.9\linewidth]{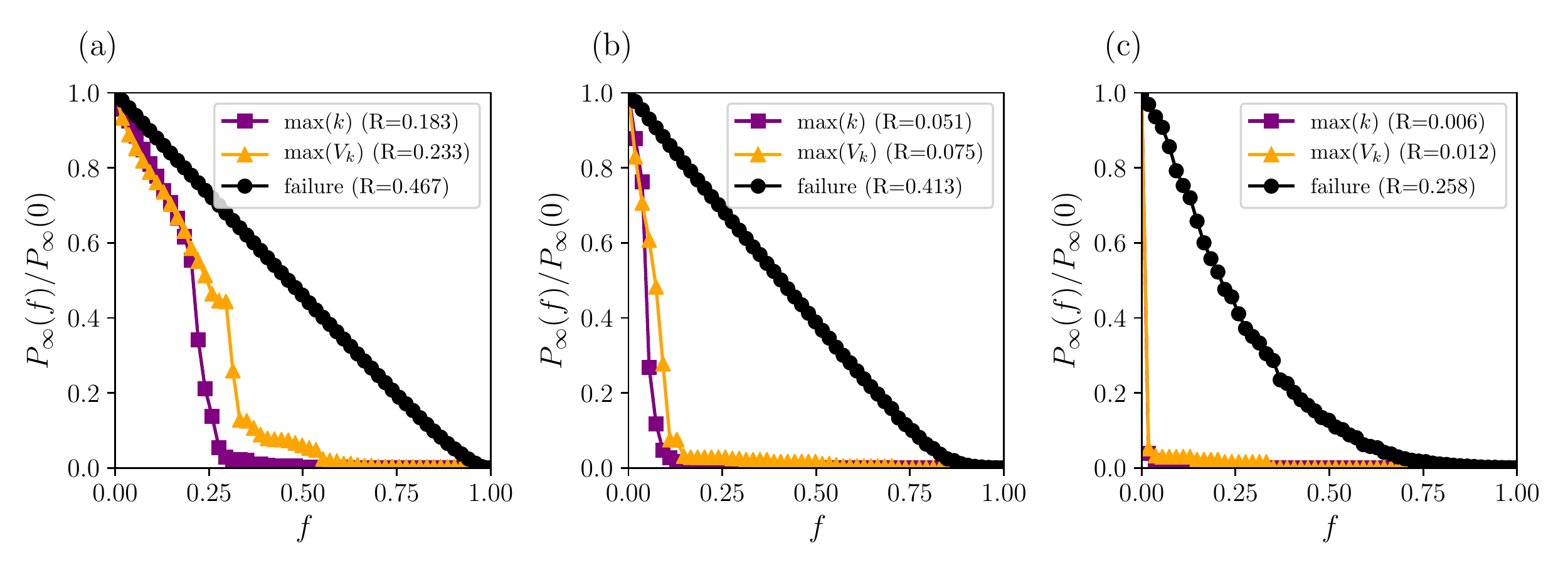}
    \caption{Robustness analysis for the Brazilian mobility network (BR), considering three thresholds: a) $\eta_1$; b) $\eta_2$; and c) $\eta_3$, as in Table \ref{tab:networks}. 
    It exhibits the rate of removed nodes $f$ versus the normalized size of the giant component $P_\infty(f) / P_\infty(0)$. 
    }
    \label{fig:robustness_brazil}
\end{figure}

Figure \ref{fig:robustness_brazil} exhibits the results for robustness\footnote{Source code available at \url{https://github.com/vanderfreitas/mobility_IBGE_2016}} in the Brazilian network under thresholds $\eta_1$, $\eta_2$ and $\eta_3$. 
It presents the size of the giant component for different node removal strategies. 
The lefthand picture corresponds to the network built directly from the original data ($\eta_1$), and the righthand one is the least connected ($\eta_3$).

As expected \cite{NetScienceB}, random failures do not break the network until almost all nodes are removed, due to its scale-free structure. The targeted attacks in nodes with higher $V_k$ ($\max(V_k)$) dismantle the networks for small $f$ as the attacks with higher degrees ($\max(k)$). The higher the threshold, the fewer nodes must be removed to break the network structure, since it shortens the number of connections and increases both the nodes' degrees and the shortest paths.  



\begin{figure}[H]
    \centering
    \includegraphics[width=0.9\linewidth]{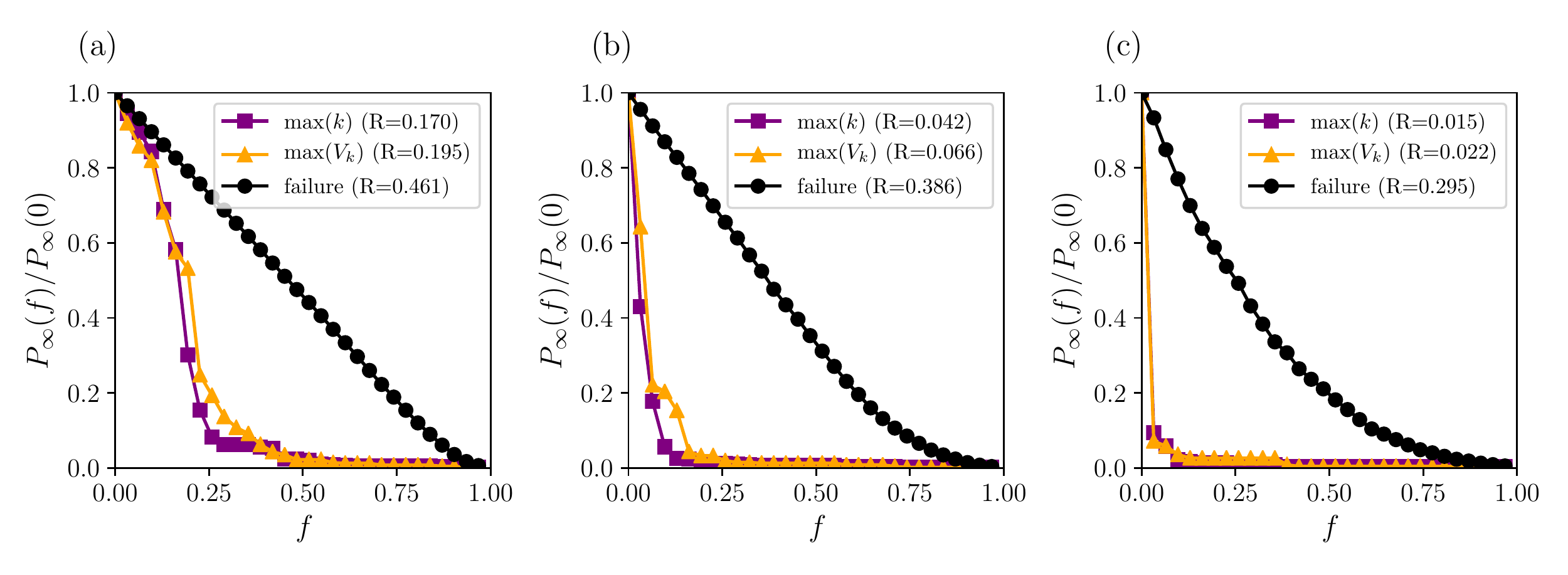}
    \caption{Robustness analysis for the São Paulo mobility network (SP), considering three thresholds: a) $\eta_1$; b) $\eta_2$; and c) $\eta_3$, as in Table \ref{tab:networks}. 
    It exhibits the rate of removed nodes $f$ versus the normalized size of the giant component $P_\infty(f) / P_\infty(0)$. 
    }
    \label{fig:robustness_SP_state}
\end{figure}

The São Paulo mobility network (SP) produces similar results as the BR's (Figure \ref{fig:robustness_SP_state}). The main difference is that it takes longer to break, since, although it has fewer nodes and edges, it is more connected, with an average degree 28.5\% higher. 

\begin{figure}[H]
    \centering
    \includegraphics[width=0.9\linewidth]{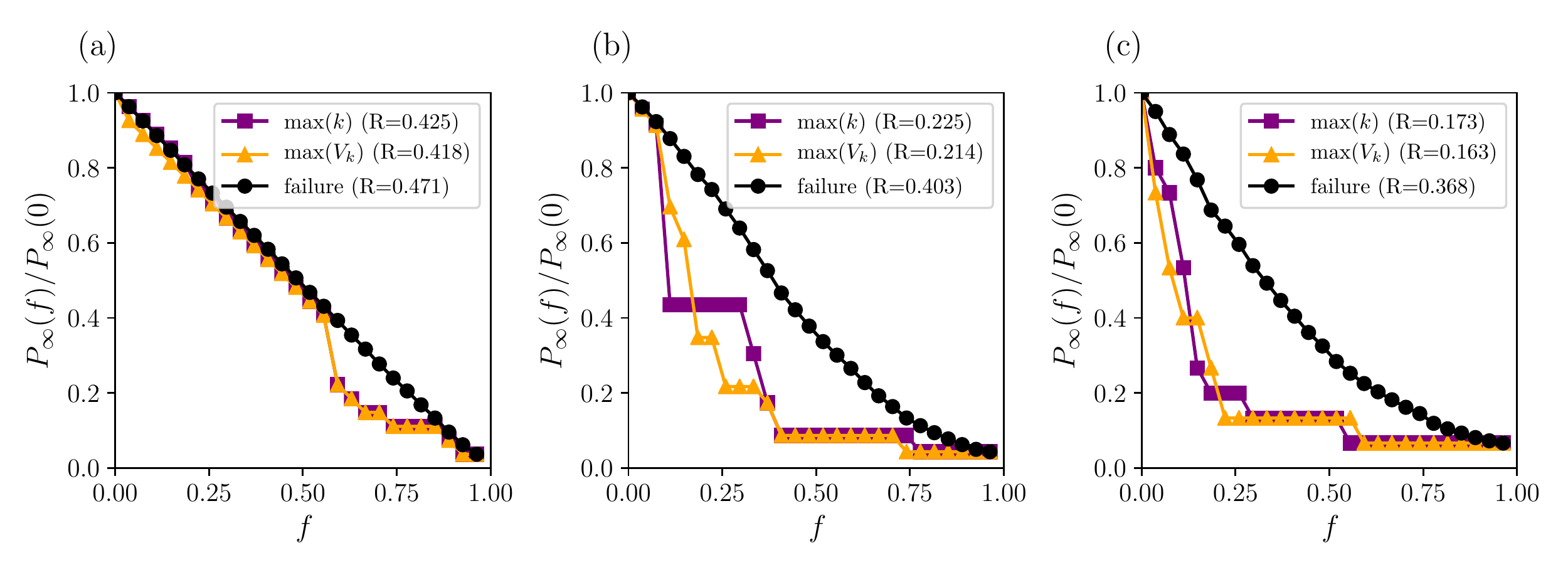}
    \caption{Robustness analysis for the Brazilian states' mobility network (BS), considering three thresholds: a) $\eta_1$; b) $\eta_2$; and c) $\eta_3$, as in Table \ref{tab:networks}. 
    It exhibits the rate of removed nodes $f$ versus the normalized size of the giant component $P_\infty(f) / P_\infty(0)$. 
    }
    \label{fig:robustness_brazilian_states}
\end{figure}

The differences between failures and attacks are only noticeable for higher thresholds in the network formed by the Brazilian states (BS) (Figure \ref{fig:robustness_brazilian_states}). The network for $\eta_1$ is not scale-free but also gives similar results for vulnerability and connectivity. Removing nodes with higher $k$ or $V_k$ does not cause more impact than picking by chance. The results become to differ for other thresholds when the shortest paths between nodes increase. Note that there are some plateaus that represent regions where the removal of some nodes does not impact on robustness (refer, for example, to the interval $f \in [0.2, 0.55]$ in Figure \ref{fig:robustness_brazilian_states}c). 

There is one interesting transition from the BR network, passing by the SP and reaching the BS, regarding the two attack strategies. The $\max(V_k)$ for $\eta_1$ is less effective than $\max(k)$ in BR, gives nearly the same results in SP and is better than $\max(k)$ in BS.

The city of São Paulo is in the first position and the city of Belo Horizonte is in the top list in all cases: all thresholds and both $\max(V_k)$ and $\max(k)$. For low threshold values, the city of Feira de Santana (state of Bahia) appears in the top list while the city of Salvador (capital of Bahia) does not, probably because it is the second-largest city of the state and connects the capital do the countryside. Besides, it is the second-largest countryside Brazilian city  in the North, Northeast, East, and South of the country\footnote{Source (IBGE): \url{https://agenciadenoticias.ibge.gov.br/agencia-sala-de-imprensa/2013-agencia-de-noticias/releases/25278-ibge-divulga-as-estimativas-da-populacao-dos-municipios-para-2019}}.  For high threshold values, the city of Campina Grande (state of Paraíba) also appears in the top list, while the city of João Pessoa (capital of Paraíba) does not. Campina Grande, alike Feira de Santana, possesses one of the most important bus transport systems in the Northeast region and connects many cities to the sea coast. Note that within the context of an epidemic, such cities are potential super spreaders along with the states' capitals.

The presented robustness analysis can be applied in different domains as well, from infectious diseases, as the main motivation for this paper due to the Corona Virus outbreak in Brazil, to the mitigation of extreme events and natural hazards into critical infrastructure networks, such as wildfires in power grids and floods and landslides in highways. Such mobility-based analysis is important for urban planning as well, for regional development, especially in a continental dimension country like Brazil.


\section{Final remarks}
\label{sec:final_remarks}

We present a robustness analysis into inter-cities mobility complex networks with the abstraction of municipal initiatives as nodes' failures and the federal actions as targeted attacks. The networks are built with the IBGE mobility data on roads and waterways from 2016. It accounts for the flow of buses, vans and aquatic transports between cities, considering only vehicles from companies that sell tickets to passengers. 

Cities (or states) are modeled as nodes in the network and the connections are mediated through the mobility data. The isolation of certain nodes is extremely relevant to spreading process containment. The question we address in the paper is to determine what are the most important nodes that keep the network connected. We consider three scenarios: the whole network with the $N=5420$ Brazilian cities, another with the cities of the São Paulo state only ($N=620$) and lastly the network formed by the $N=27$ Brazilian states, each as a node.

The abstraction we make is to consider the random removal of nodes (failures) as city initiatives, that do not have a connection to the country policies. Conversely, the attacks are isolation measures determined by the federal government. Such attacks are performed according to some node indexes like degree and Vulnerability, the latter as a measure of how vulnerable the network is in the node absence.

The indexes we use do not account for the flow information, so we performed an analysis of each network in three instances, considering different thresholds for the flow. The first instance comprehends all links with nonzero flows, the second contains edges with flows above a certain average value, and lastly the flows above a high threshold. The instances capture decreasing mobility due to pandemics and less important routes closed by the companies.

Our results show that the federal actions have a strong impact on the network, while the local ones do not break it. Choosing the cities with higher degrees for the targeted attacks is the best option in most cases, especially for the larger network of $N=5420$ cities. However, there is a transition, showing that the Vulnerability index performs nearly the same as the degree for the São Paulo network and is the best choice for the smallest (states).

Moreover, the higher the threshold the more effective the attacks are, due to the increasing smaller path lengths between nodes when computing Vulnerability, and also because the number of less connected nodes decreases and the remaining are mostly hubs. Failures also impact more, but the transition is not as significant as for attacks.

Surprisingly, some countryside cities such as Campina Grande (state of Paraíba) and Feira de Santana (state of Bahia) have higher degrees and Vulnerability indexes than some states' capitals. Their importance in mobility is crucial and their absence leads to the network collapse.

In future works, we hope to perform attacks with indexes that explicitly take the flows into account (weighted approaches). Besides, that would be interesting to perform simulations with epidemic models such as the Branching Processes and variations of SIR (susceptible, infected and removed) \cite{NetScienceB, Cota} on top of the investigated networks to assess how the attacks help stop the spreading processes. 






\section{Funding}
\label{sec:agrad}
São Paulo Research Foundation (FAPESP), Grant Number 2015/50122-0 and DFG-IRTG Grant Number 1740/2; FAPESP Grant Number 2018/06205-7; CNPq Grant Number 420338/2018-7

\bibliography{mybibfile}

\end{document}